\documentclass[12pt]{article}

\usepackage{newtxtext,newtxmath}

\usepackage{graphicx}

\usepackage{subcaption}
\usepackage{subfig}
\usepackage{lmodern}

\usepackage[letterpaper,margin=1in]{geometry}

\renewenvironment{abstract}
	{\quotation}
	{\endquotation}

\date{}

\makeatletter
\renewcommand{\fnum@figure}{\textbf{Figure \thefigure}}
\renewcommand{\fnum@table}{\textbf{Table \thetable}}
\makeatother

\usepackage{scicite}

\usepackage{url}

\def\scititle{
	Slovakia's Mass Testing: A Critical Look at the Negative Effects
}
\title{\bfseries \boldmath \scititle}

\author{
	Jozef \v{C}ern\'{a}k
	\\
	\small Diamantov\'{a} 8, Ko\v{s}ice, SK-04011, Slovak Republic,
	\small Email: jozefcernak@gmail.com\and
}

\begin{document} 

\maketitle

\begin{abstract} \bfseries \boldmath
This e-letter re-evaluates the epidemiological impact of nationwide mass antigen testing in Slovakia. While initial reports \cite{Pavelka} proposed a causal link between these campaigns and declining viral prevalence, granular re-analysis reveals a significant temporal mismatch. We argue that the proclaimed success represents a conceptual nexus lacking empirical support; shifts in the effective reproduction number ($R_t$), case trajectories, and mortality rates do not align with the testing rounds. Crucially, the mortality-to-hospital admission ratio exhibits a distinct inverse relationship with the interventions. Rather than providing a clinical benefit, the testing campaigns were followed by increased mortality and a strained healthcare system. We contend that these adverse outcomes were a direct consequence of the testing policy, which sustained higher overall mobility levels compared to the United Kingdom. By overattributing causality to mass testing, a spurious nexus was constructed, obscuring the true drivers of the pandemic and its socio-economic consequences.
\end{abstract}

\noindent

For the intervention \cite{Pavelka} to be considered effective, its impact should manifest as a discernible shift in the reproduction number $R_t$ \cite{Flaxman2020, Soltesz2020, Flaxman2020_R, Contreras, Linka2020, Angeli2025}, new case trajectories, mortality rates \cite{Prada2022}, and the mortality-to-hospital admission ratio \cite{Cernak_2026}. However, a granular analysis of temporal trends reveals that shifts in $R_t$, as well as the trajectories of new cases (Fig. \ref{fig:Fig_S1} in \cite{Cernak_2026}) and deaths (Fig. \ref{fig:Fig_S2} in \cite{Cernak_2026}), do not exhibit positive trends associated with the timing of the testing campaigns. Consequently, we demonstrate the absence of any identifiable beneficial effect on $R_t$ (Figs. \ref{fig:Fig_S1}, \ref{fig:Fig_S7} in \cite{Cernak_2026}), case trends, mortality (Fig. \ref{fig:Fig_S2} in \cite{Cernak_2026}), or the mortality-to-hospital admission ratio (Fig. \ref{fig:Fig_S3} in \cite{Cernak_2026}) across both short- and long-term horizons.

To demonstrate the positive impact of mass testing, Pavelka et al. \cite{Pavelka} claimed: ``By contrast, data on hospital bed occupancy shows a sudden flattening from mid-November, indicating a sharp decrease in new admissions that is primarily consistent with a sizable reduction in new infections when the mass testing campaigns occurred (fig. S6).'' We consider this interpretation to be unsubstantiated.

We observe that as a pandemic wave subsides—as evidenced in neighboring regions—the incidence of new hospital admissions, daily mortality, and total hospitalizations should concurrently decline (Fig. \ref{fig:Fig_S2} in \cite{Cernak_2026}; e.g., the outbreak dynamics in Czechia). Consequently, the observed stagnation or fluctuation in hospital admissions, daily deaths, and the total number of hospitalized patients in Slovakia indicates that these metrics were, in fact, elevated relative to a scenario of natural wave recession.

Furthermore, since hospital bed occupancy—as adopted by Pavelka et al. \cite{Pavelka}—is influenced by multiple confounding factors, including varying admission and discharge rates, it is an unreliable metric for assessing an intervention's success. We contend that mortality dynamics \cite{Prada2022} and more granular indicators of clinical burden \cite{Spencer_2022} provide more accurate insights into the impact of mass testing. Contrary to the reported success, we demonstrate adverse outcomes by examining the mortality-to-hospital admission ratio (Fig. \ref{fig:Fig_S3} in \cite{Cernak_2026}). This ratio increased immediately following the testing campaigns, fluctuated near its peak for $30$ days, and subsequently rose to a second, higher plateau. Notably, no such abrupt escalation was observed in Czechia, where conventional mitigation measures were maintained.

Furthermore, Pavelka et al. \cite{Pavelka} exhibit inconsistency in their comparative analysis of intervention efficacy. They claimed: ``For comparison, a month-long lockdown in November in the UK resulted in just a 30\% decrease in prevalence. This, alongside the inability in December to control the rebounding spread in Slovakia through even more stringent contact restrictions, indicates that the mass testing campaigns were responsible for a large share of case reduction in the previous months.'' This conclusion is methodologically problematic, as it draws a false equivalence between disparate epidemiological contexts without providing direct empirical evidence. By over-relying on such cross-national comparisons, the authors overlook the specific drivers of transmission and the socio-economic variables unique to each region.

The comparison by Pavelka et al. \cite{Pavelka} between the UK lockdown and Slovak mass testing is fundamentally flawed, as it overlooks the critical divergence in effective reproduction numbers and constructs a spurious nexus by isolating mass testing from the broader epidemiological context. While the UK's month-long lockdown successfully suppressed viral circulation—achieving a sustained decline in $R_t$ to approximately $0.75$—the Slovak approach failed to yield a comparable impact. Our analysis demonstrates that despite intensive testing campaigns, $R_t$ in Slovakia fluctuated around $1.0$ for an extended period, failing to break the overall momentum of the outbreak. This suggests that the perceived inability to control the spread in December was not a failure of contact restrictions per se, but rather a consequence of a testing strategy that sustained high levels of mobility and social mixing. By overattributing causality to mass testing, the authors \cite{Pavelka} obscure the true drivers of the pandemic, including the seasonal progression of the winter wave and the inherent limitations of using testing as a functional substitute for stringent movement restrictions.

Our analysis demonstrates that the observed negative trends—specifically the abrupt escalation in mortality (Fig. \ref{fig:Fig_S2} in \cite{Cernak_2026}) and the systemic exhaustion of the Slovak healthcare system (Fig. \ref{fig:Fig_S3} in \cite{Cernak_2026}) following mass testing—indicate that the conclusions of Pavelka et al. \cite{Pavelka} are based on a causal misattribution.

In conclusion, the prioritization of mass testing as a primary epidemiological strategy, which failed to reduce the reproduction number $R_t$ significantly below $1.0$, appears to have had severe public health implications. By sustaining a baseline of higher mobility and social mixing compared to the United Kingdom, this approach hindered the effective suppression of viral transmission. This ultimately contributed to a protracted surge in infections that overstretched the Slovak healthcare system, likely resulting in suboptimal care conditions during the mortality plateau of early $2021$. Our findings suggest that a substantial portion of the human cost was intrinsically linked to the strategic limitations of the chosen intervention.


\clearpage 

%
\bibliography{science_cernak} 
\bibliographystyle{sciencemag}

%
%
%
%
%
%



\paragraph*{Funding:}
The author received no specific funding for this work.
\paragraph*{Author contributions:}
 J.C. is the sole author of this work and was responsible for the conceptualization, data analysis, and writing of the comment.
\paragraph*{Competing interests:}
The author(s) declare no competing interests.
\paragraph*{Data and materials availability:}
The epidemiological and mobility data, along with the Gnuplot scripts used to generate the figures and reproduce the results in this e-letter, are publicly archived on Zenodo at \url{https://doi.org/10.5281/zenodo.19431180} \cite{Cernak_2026_data}. This repository includes processed time-series for the effective reproduction number $R_t$, mortality-to-hospital admission ratios, and mobility indices for Slovakia, the Czech Republic, and the United Kingdom (Version 1.0.0). All underlying raw data were retrieved from the Google Community Mobility Reports and official national health statistics, which are cited within the repository's metadata. There are no restrictions on the reuse of these materials under the Creative Commons Attribution 4.0 International (CC-BY 4.0) license.


\subsection*{Supplementary materials}
Materials and Methods\\
Supplementary Text\\
Figs. S1 to S7\\
Table S1 \\
References \textit{(1-\arabic{enumiv})}\\ 
Data S1 \textbf{Slovakia's Mass Testing: A Critical Look at the Negative Effects (Dataset and Figures) \cite{Cernak_2026_data}}


\newpage


\renewcommand{\thefigure}{S\arabic{figure}}
\renewcommand{\thetable}{S\arabic{table}}
\renewcommand{\theequation}{S\arabic{equation}}
\renewcommand{\thepage}{S\arabic{page}}
\setcounter{figure}{0}
\setcounter{table}{0}
\setcounter{equation}{0}
\setcounter{page}{1} 


\begin{center}
\section*{Supplementary Materials}

Jozef \v{C}ern\'{a}k,\\
\small Diamantov\'{a} 8, Kosice, SK-04011, Slovak Republic
\small Email: jozefcernak@gmail.com\\

\end{center}

\subsubsection*{This PDF file includes:}
Materials and Methods\\
Supplementary Text\\
Figures S1 to S7\\
Table S1 \\
Captions for Data S1\\
References \textit{(1-\arabic{enumiv})}\\
\subsubsection*{Other Supplementary Materials for this manuscript:}
Data S1 \textbf{Slovakia's Mass Testing: A Critical Look at the Negative Effects (Dataset and Figures)}

\newpage


\subsection*{Materials and Methods}

The conclusions drawn by Pavelka et al. \cite{Pavelka} were primarily derived from national COVID-19 statistics spanning mid-September to late December 2020, supplemented by an assessment of the pandemic context in Slovakia and the UK. In contrast, our study utilized open-access data from the Our World in Data repository \cite{owid_coronavirus} and expanded the geographical scope to include the Czech Republic. This addition is critical as the Czech Republic shared a highly synchronized epidemiological trajectory with Slovakia until the tipping point in early November 2020 during the second wave \cite{Cernak}, yet implemented divergent non-pharmaceutical intervention (NPI) strategies.

We extended the analytical timeframe by incorporating data up to February 22, 2021. This cutoff date was specifically selected to encompass the peak of the winter wave in both countries, while ensuring the inclusion of data that were already available to the authors of the original study \cite{Pavelka} prior to their manuscript’s final publication. This standardized window facilitates a robust comparative analysis of how varying national measures influenced viral transmission dynamics across two closely related Central European contexts.

\subsubsection*{Comparative Approach}
Our study adopts the comparative framework established by Pavelka et al. \cite{Pavelka}. To investigate the potential decoupling between mass testing interventions and their subsequent short- and long-term impacts, we shifted the analytical focus toward the Czech Republic—an approach consistent with our previous research \cite{Cernak}. Furthermore, we incorporated key parameters of the pandemic dynamics from the United Kingdom, specifically mobility patterns (Fig. \ref{fig:Fig_S6}) and the reproduction number $R_t$ dynamics (Fig.\ref {fig:Fig_S7}), to establish a broader contextual baseline for our findings. This dual comparison allows for a more rigorous evaluation of whether the observed shifts in epidemiological indicators can be genuinely attributed to mass testing or to broader regional trends.

\subsubsection*{Considering additional parameers}

We evaluate the mortality-to-hospital admission ratio ($MHR$), defined as:

\begin{equation}
	MHR = \frac{D}{A} \times 100
	\label{eq:sup_MHR}
\end{equation}

where $D$ denotes the number of deaths and $A$ denotes the number of hospital admissions within a specified time period. In our analysis, we utilize a one-day interval. Given the observed temporal dynamics of admissions, deaths, and hospital occupancy (Fig. \ref{fig:Fig_S2}), the robustness of our results (Fig. \ref{fig:Fig_S3}) remains intact; even with a single-day resolution, the findings are not significantly altered despite the inherent time lag in mortality.

Unlike raw mortality counts (Fig. \ref{fig:Fig_S2}) or continuously growing hospital admissions (Fig. \ref{fig:Fig_S2}), the $MHR$ accounts for the varying volume of patients entering or being discharged from the healthcare system. This provides a more accurate normalization for the severity of the outbreak than the metrics used in the initial study \cite{Pavelka}. Furthermore, the $MHR$ serves as a critical proxy for the clinical burden; elevated $MHR$ values may indicate periods of acute healthcare strain, where limited capacity and staff shortages potentially compromise patient outcomes. Consequently, this ratio reflects the relative mortality risk within the hospitalized population under varying degrees of systemic pressure.

\subsubsection*{Legislative frameworks for mandatory testing in Slovakia}
The implementation of mandatory testing was supported by a series of legal measures. The following key regulations established the framework for population-wide testing (original titles in Slovak):

\begin{enumerate}
  \item UZNESENIE VL\'{A}DY SLOVENSKEJ REPUBLIKY
č. 693 z 28. okt\'{o}bra 2020 k n\'{a}vrhu na \v{d}al\v{s}ie roz\v{s}írenie opatren\'{i} v r\'{a}mci vyhl\'{a}sen\'{e}ho n\'{u}dzov\'{e}ho stavu pod\v{l}a \v{c}l. 5 \'{u}stavného zákona č. 227/2002 Z. z. o bezpečnosti štátu v čase vojny, vojnového stavu, výnimočného stavu a núdzového stavu v znení neskorších predpisov vyhláseného uznesením vlády Slovenskej republiky č. 587 z 30. septembra 2020

  \item The numbers start Vestník vlády Slovenskej republiky, Ročník 30, Čiastka 12, Vydaná 30. októbra 2020, 16
VYHLÁŠKA Úradu verejného zdravotníctva Slovenskej republiky, ktorou sa nariaďujú opatrenia pri ohrození verejného zdravia k režimu vstupu osôb do priestorov prevádzok a priestorov zamestnávateľa

  \item Vestník vlády Slovenskej republiky, Ročník 31, Čiastka 19, Vydaná 5. februára 2021, 47 VYHLÁŠKA Úradu verejného zdravotníctva Slovenskej republiky, ktorou sa nariaďujú opatrenia pri ohrození verejného zdravia k režimu vstupu osôb do priestorov prevádzok a priestorov zamestnávateľa
\end{enumerate}

These regulations aimed to pressure citizens into getting tested by linking testing to mobility rights, such as permission to go to work or attend social events from which the 'untested' were excluded.


\subsection*{Supplementary Text}

\subsubsection*{Dynamics of outbreak waves: A comparison of Slovakia and the Czech Republic}

\begin{figure} 
	\centering
	\includegraphics[width=0.6\textwidth]{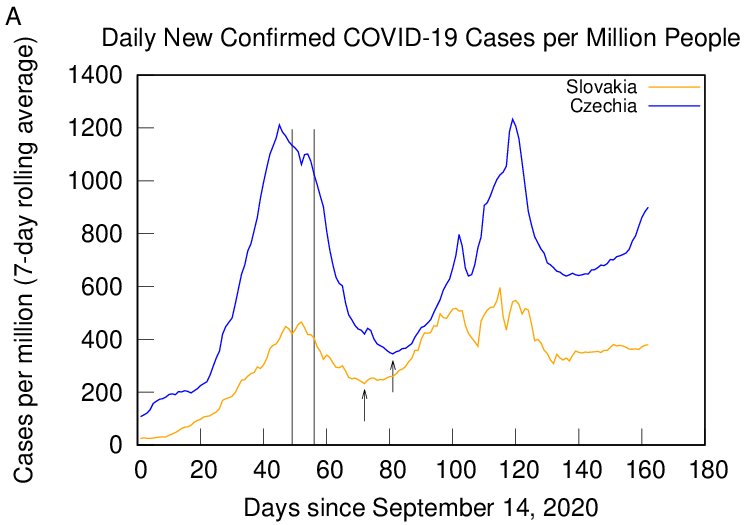} 
	\includegraphics[width=0.6\textwidth]{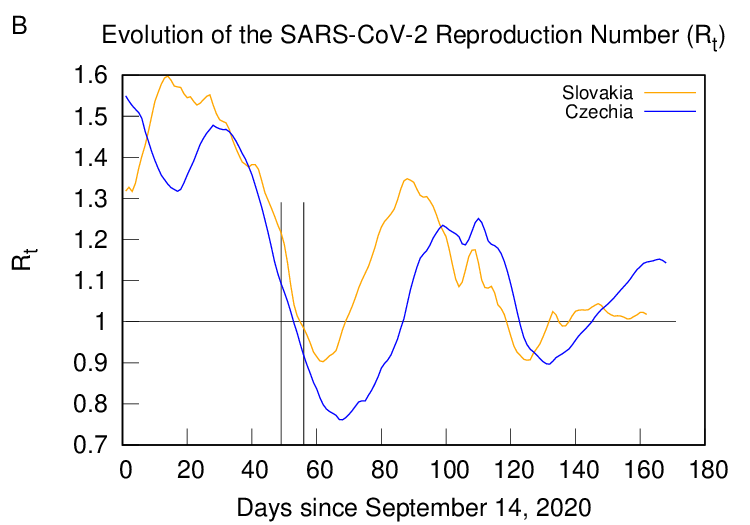}


	\caption{\textbf{Dynamics of new cases and effective reproduction number $R_t$ in Slovakia and Czechia (September 2020–February 2021).}
Vertical lines denote mass testing events in Slovakia. The top panel (\textbf{A}) shows daily new confirmed COVID-19 cases per million people. The bottom panel (\textbf{B}) reveals the evolution of the SARS-CoV-2 reproduction number $R_t$.}

	\label{fig:Fig_S1} 
\end{figure}

Slovakia and the Czech Republic share a profound historical and geographical proximity, which was reflected in their highly synchronized epidemiological trends during the autumn of 2020, prior to the implementation of mass testing in Slovakia \cite{Cernak}. As shown in Figure \ref{fig:Fig_S1}, the growth phases of the outbreak waves and the trajectories of the effective reproduction number ($R_t$) followed nearly identical paths in both nations. However, the respective peaks of these waves marked a critical turning point; while the initial dynamics were shared, the subsequent divergence after these peaks reveals substantial differences in the effectiveness of the pandemic responses in each country.

In Slovakia, this turning point coincided with the implementation of the nationwide mass testing campaign (Fig. \ref{fig:Fig_S1}). Our analysis demonstrates that this intervention disrupted the previously synchronized epidemiological patterns between the two countries (Fig. \ref{fig:Fig_S1}). Contrary to the anticipated improvement, the situation in Slovakia failed to stabilize, and the natural alignment with the Czech trajectory was severed. Following the peak, the dynamics of new cases and the reproduction number in Slovakia did not exhibit a standard decline. This indicates that mass testing failed to deliver the expected epidemiological breakthrough, leading instead to a divergent and non-improving trend.

Our data analysis (Fig. \ref{fig:Fig_S1}) identifies a significant temporal shift in the outbreak waves between the two countries. While the peak of the autumn wave in Slovakia lagged approximately seven days behind that of the Czech Republic, the onset of the subsequent winter wave occurred ten days earlier in Slovakia. Notably, this accelerated onset of the winter wave began almost immediately following the nationwide mass testing campaign. This suggests that the intervention failed to provide the anticipated epidemiological respite; instead, the synchronization between the two nations was disrupted by an earlier and more aggressive resurgence in Slovakia, leading to premature strain on the healthcare system. This advanced onset of the winter wave is directly corroborated by the observed dynamics of the effective reproduction number $R_t$ (Fig. \ref{fig:Fig_S1}).

In Slovakia, the nationwide mass testing campaign resulted in only a transient epidemiological improvement. Our data (Fig. \ref{fig:Fig_S1}) show  that while the effective reproduction number briefly dropped below 1.0, this effect was limited to approximately 14 days, reaching a minimum of $R_t = 0.90$. In sharp contrast, the decline in the Czech Republic during the same period was significantly more pronounced ($R_t = 0.76$) and sustained, with $R_t$ remaining below 1.0 for nearly 33 days. Following this short-lived and marginal dip, $R_t$ in Slovakia quickly returned to values above 1.0, failing to break the overall momentum of the outbreak. Consequently, the situation continued to deteriorate, leading to a systemic strain approximately 30 days after the testing campaign, as hospitalizations reached a critical threshold of approximately 1700 patients—a capacity limit significantly lower than officially projected.

\begin{figure} 
	\centering
	\includegraphics[width=0.6\textwidth]{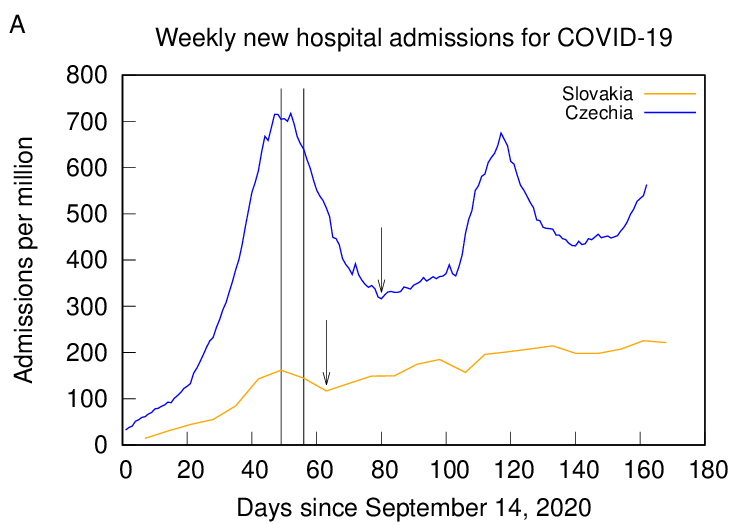} 
	\includegraphics[width=0.6\textwidth]{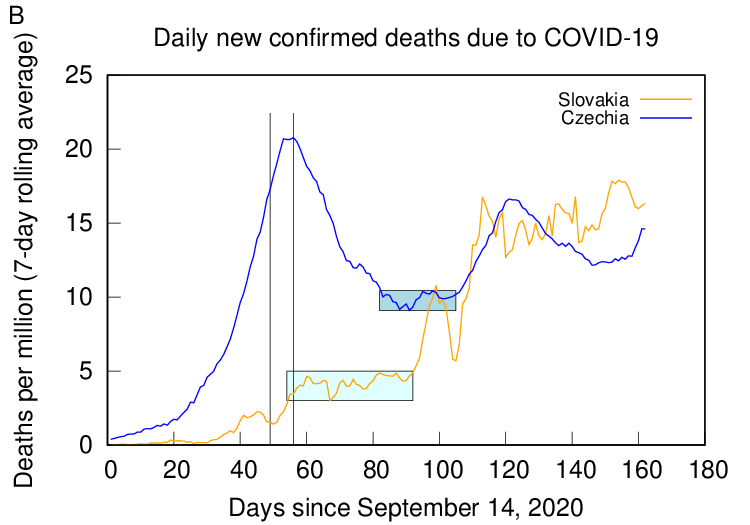}
	\includegraphics[width=0.6\textwidth]{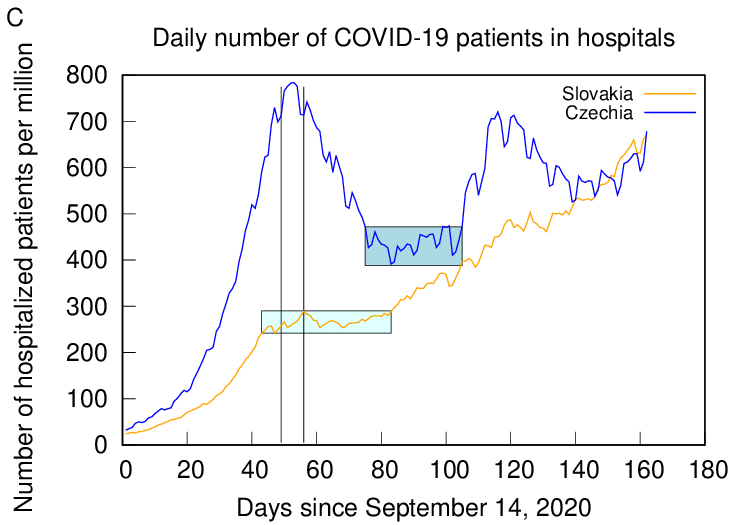}

\caption{\textbf{ }
caption next page
}	
	
\label{fig:Fig_S2} 

\end{figure}

Following the nationwide mass testing and the conclusions of Pavelka et al. \cite{Pavelka}, it would have been logical to expect a discernible decrease in weekly hospital admissions, daily new confirmed deaths, and the total number of hospitalized patients; however, such trends were not empirically observed. Instead, our analysis (Fig. \ref{fig:Fig_S2}) reveals only a marginal and transient deceleration in weekly new hospital admissions compared to the trajectories in the Czech Republic. This was followed by a subsequent sustained increase that commenced 17 days earlier than the corresponding trend in the Czech Republic.

While new confirmed deaths in the Czech Republic declined dramatically, mortality in Slovakia continued to fluctuate for 38 days at the peak levels established during the mass testing campaign (Fig. \ref{fig:Fig_S2}). Following this protracted stagnation, the number of confirmed deaths began to rise sharply, reaching significantly higher levels that remained sustained over the long term. This indicates that the intervention failed to achieve a meaningful reduction in mortality, in sharp contrast to the effective suppression observed in the Czech Republic.

Similar adverse trends were observed in the daily census of hospitalized COVID-19 patients (Fig. \ref{fig:Fig_S2}). In Slovakia, this number fluctuated for 40 days around the peak values recorded during the mass testing campaign, followed by a subsequent sustained and long-term increase.

In summary, the data (Fig. \ref{fig:Fig_S2}) demonstrate that the expected short-term epidemiological reprieve never materialized. Instead, the brief stabilization of key metrics was immediately followed by profound long-term adverse trends, characterized by a premature winter wave and a sustained escalation in mortality. This divergence from the Czech trajectory suggests a systemic failure, further evidenced in the mortality-to-hospital admission ratio (Fig. \ref{fig:Fig_S3}), which illustrates the subsequent deterioration in the quality of clinical care and hospital outcomes.

Based on the aforementioned results (Fig. \ref{fig:Fig_S2}), the healthcare system in Slovakia was subjected to sustained pressure without the opportunity to decompress during a declining outbreak wave, as was observed in the Czech Republic. We contend that the systemic strain on the Slovak healthcare system, manifesting approximately 30 days after the mass testing campaign (Fig. \ref{fig:Fig_S3}), represents a significant long-term adverse impact that Pavelka et al. \cite{Pavelka} failed to account for in their analysis.

\subsubsection*{Clinical Care Quality and Hospital Outcomes}

The systemic failure described above is further elucidated in Figure \ref{fig:Fig_S3}, which depicts the relationship between hospital occupancy and clinical outcomes. Our analysis reveals a critical threshold reached on December 12, 2020, beyond which the quality of care significantly deteriorated.

Unlike the Czech Republic, where the healthcare system experienced a period of relief that facilitated staff recovery and resource replenishment, the Slovak hospital network remained under continuous, high-intensity strain. This absence of a 'decompression phase' led to a progressive exhaustion of both human and material resources. Consequently, as Figures \ref{fig:Fig_S2} and \ref{fig:Fig_S3} demonstrate, confirmed deaths and the number of hospitalized patients began to rise disproportionately to the total number of admissions. This trend indicates that the failure was not merely a matter of absolute bed capacity, but a degradation of care quality under prolonged systemic stress. The data suggest that patients admitted during the post-testing period faced a higher probability of adverse outcomes compared to those in the initial phase of the wave, highlighting the severe long-term consequences of failing to suppress the momentum of the outbreak in November 2020.

Following the nationwide mass testing campaign, the volume of tests increased continuously for 65 days, culminating in a dramatic surge at the beginning of 2021 (Fig. \ref{fig:Fig_S3}), when a further abrupt escalation in testing occurred. These facts demonstrate that mass testing, even when followed by a sustained expansion of testing capacity, failed to prevent the systemic collapse of the Slovak healthcare system.

\subsubsection*{The Role of Mobility and Behavioral Response 35 days after mass testing}

The divergent epidemiological trajectories between Slovakia and the Czech Republic cannot be explained by simple differences in overall mobility volume. Data from Google Community Mobility Reports Figure \ref{fig:Fig_S5} indicate that for approximately 35 days following the mass testing campaign, mobility trends related to workplaces and public transport remained remarkably similar in both countries. The only significant divergence was observed in outdoor and park mobility, which was notably higher in Slovakia—a trend that continued even as the epidemic worsened.

However, a broader international comparison reveals a more fundamental failure in the Slovak strategy. When compared to the United Kingdom during the corresponding 35-day post-intervention period (Fig. \ref{fig:Fig_S5}), Slovakia exhibited significantly higher mobility levels across retail and recreation, public transport, and workplaces. While the UK implemented stringent movement controls that effectively suppressed viral circulation, the Slovak approach—relying on mass testing as a functional substitute for a rigorous lockdown—permitted sustained high-frequency interactions in high-risk environments.

This lack of effective movement restrictions is directly reflected in the divergence of epidemiological indicators. While the effective reproduction numbers ($R_t$) in both the Czech Republic and the United Kingdom exhibited a sustained decline during this period, the trajectory in Slovakia was markedly different. Although $R_t$ in Slovakia had initially been decreasing, it began to rise almost immediately following the mass testing campaign, remaining elevated and failing to break the overall momentum of the outbreak.

In summary, the strategic failure in Slovakia was two-fold. First, there was a failure to account for the risk-compensation behavior triggered by the initial mass testing results and the subsequent strategy of permanent population testing, which collectively fostered a sustained false sense of security. Second, the administration failed to implement the requisite level of movement restrictions, as evidenced by higher mobility levels compared to both the Czech Republic (in the outdoor sector) and the United Kingdom (across all sectors). This combination inadvertently maintained viral circulation and denied the Slovak healthcare system the critical 'decompression phase' achieved by its neighbors.

\subsubsection*{Mobility patterns at the beginning of 2021}

Data from the Google Community Mobility Reports (Figs. \ref{fig:Fig_S5}, \ref{fig:Fig_S6}) indicate that while mobility in the retail and recreation sectors remained suppressed across the United Kingdom, the Czech Republic, and Slovakia, mobility within workplaces and public transport hubs exhibited a high degree of resilience. The most pronounced reductions in public transport and workplace mobility were recorded in the United Kingdom, whereas significantly higher levels were observed in Slovakia, with the highest relative mobility rates found in the Czech Republic.

\subsubsection*{Sustained $R_t \approx 1$ Conditions at the beginning of 2021}

As a consequence of the more stringent movement restrictions, the reproduction number $R_t$ in the United Kingdom dropped to approximately 0.75 and remained suppressed for an extended period. In contrast, in Czechia, the reproduction number rose from 0.92 to 1.15. The epidemiological situation in Slovakia was unique, characterized by $R_t$ stagnating slightly above 1.0 for a prolonged duration.

Despite the exceptionally high volume of COVID-19 tests performed in Slovakia during this period relative to Czechia or the United Kingdom, there was no measurable impact on maintaining $R_t$ below the critical threshold of 1.0. Notably, this lack of suppression occurred even though individuals testing positive and their close contacts were subject to a 10-day quarantine, a protocol identical to the one implemented during the nationwide mass-testing campaigns in November 2020.

In Slovakia, intensive mandatory mass testing was implemented during this period—a measure unparalleled in either the United Kingdom or the Czech Republic. The fact that the reproduction number in Slovakia stagnated around 1.0 for an extended duration, whereas it consistently decreased in the United Kingdom (Fig. \ref{fig:Fig_S7}), suggests that mandatory mass testing was an insufficiently effective tool for curbing viral transmission. This becomes particularly evident when comparing the mobility profiles: while the United Kingdom maintained the lowest mobility levels, the Czech Republic exhibited higher mobility than Slovakia.

Furthermore, the introduction of mandatory weekly testing for employees created a paradoxical mobility pattern. Instead of reducing social interactions, the policy compelled millions of citizens to regularly congregate at testing sites, maintaining a baseline of social mixing that likely neutralized the intended effects of Slovakia’s moderate mobility restrictions. In contrast to the Czech Republic, where higher mobility was eventually offset by a stabilization of healthcare metrics, the persistently higher mobility in Slovakia relative to the United Kingdom—driven by the requirement of a negative test for workplace access—contributed to a prolonged plateau of high mortality and hospital occupancy throughout the first quarter of 2021.

The phenomenon of reproduction number fluctuations and periods of stagnation near unity was recorded during various phases of the SARS-CoV-2 outbreak, as summarized in Table \ref{tab:sup_rstable}. While the underlying causes may vary, such a trend is particularly undesirable when new case counts are high, as it signalizes the ineffectiveness of measures (Fig. \ref{fig:Fig_S7}) intended to suppress viral circulation.

Such prolonged stagnation in developed countries like Slovakia (Fig. \ref{fig:Fig_S7}) demonstrates the ineffectiveness of the implemented measures—specifically, in the Slovak context, the intensive nationwide mass testing campaigns.

\section*{Conclusion}

In conclusion, our comparative analysis demonstrates that the Slovak nationwide mass testing campaign failed to achieve its primary epidemiological objectives. In contrast to the sustained suppression observed in the Czech Republic and particularly in the United Kingdom, Slovakia experienced only a transient stabilization followed by a catastrophic resurgence. This strategic failure was rooted in the inherent limitations of mass testing: it failed to provide a ``decompression phase'' for hospitals, fostered a pervasive false sense of security, and maintained high mobility levels by compelling millions to congregate for testing to retain workplace access.

Ultimately, the reliance on mass testing as a functional substitute for stringent movement restrictions led to a systemic collapse. This was characterized by a protracted strain on hospital infrastructure and a sharp, sustained escalation in mortality that was notably absent in the neighboring Czech trajectory. Our findings indicate that this strategy, by its very design, maintained the viral circulation it sought to suppress, resulting in a substantial number of preventable deaths. We contend that a significant portion of this human cost was not an inevitable consequence of the pandemic, but a direct result of a failed intervention that precluded the timely implementation of effective mitigation measures.

\addtocounter{figure}{-1}
\begin{figure} [t!]
  \caption{\textbf{Dynamics of Hospitalizations, Outcomes, and Capacity in Slovakia and Czechia (September 2020 – February 2021).}
The graphs provide a comprehensive view of the pandemic's burden on healthcare systems. The top panel (\textbf{A}) shows the weekly influx of new patients (admissions per million). The middle panel (\textbf{B}) tracks the ultimate outcomes for the most severe cases (daily deaths per million). The bottom panel (\textbf{C}) reveals the resulting strain on hospital capacity (number of patients per million). Together, they illustrate the critical balance between patient inflow (admissions) and outflow (discharges and deaths) that determines the pressure on the healthcare system. Vertical lines denote mass testing events in Slovakia. Arrows highlight the local minima, indicating transitions between pandemic waves. Shaded areas mark periods of small fluctuations or persistent shifts from the baseline.
}	
\end{figure}


\begin{figure} 
	\centering
	\includegraphics[width=0.6\textwidth]{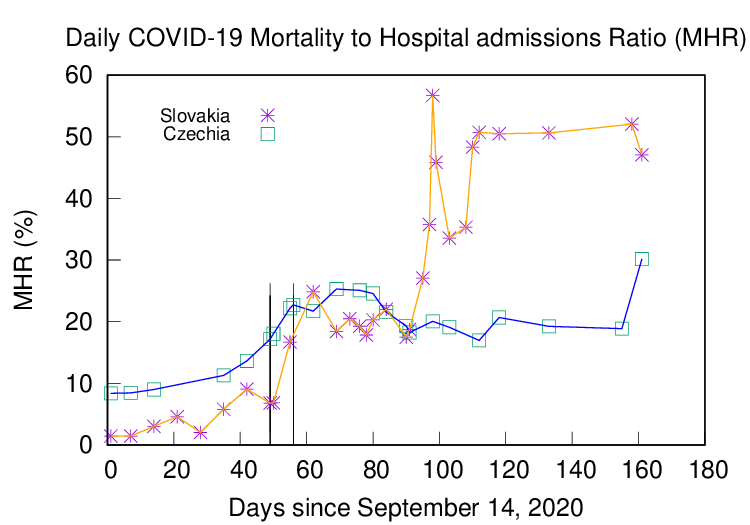} 

	\caption{\textbf{Comparison of Healthcare System Responses to COVID-19: Slovakia and Czechia (September 2020 – February 2021).}
The graph depicts the Mortality to Hospital Admission  Ratio (MHR), a key indicator of healthcare system performance. A higher ratio suggests a greater proportion of hospitalized patients did not survive, potentially indicating system overload or strain. The Czech healthcare system maintained stable performance over an extended period. In contrast, the situation in Slovakia deteriorated significantly from approximately day 90 (December 12, 2020), pointing to severe strain and a diminished capacity to provide effective care during this pandemic wave. Vertical lines denote the dates of nationwide mass testing events in Slovakia.}
	\label{fig:Fig_S3} 
\end{figure}


\begin{figure} 
	\centering
	\includegraphics[width=0.6\textwidth]{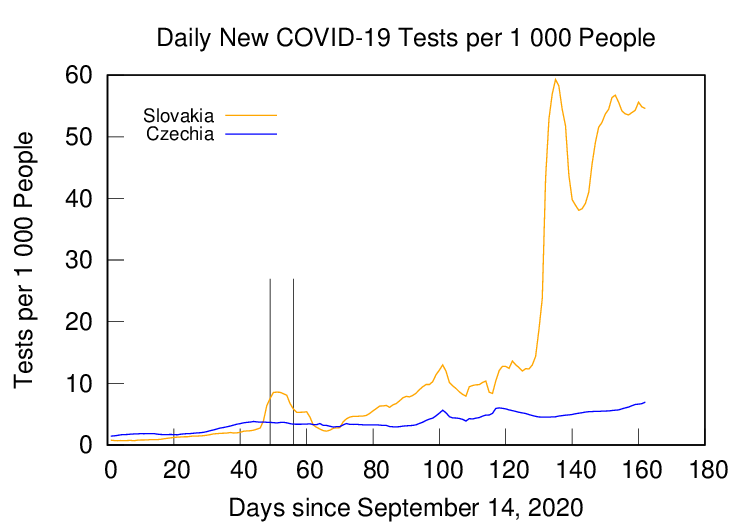} 

	\caption{\textbf{Differences in Testing Policies: Slovakia and Czechia (September 2020 – February 2021).}
The graph illustrates the divergence in COVID-19 testing intensity between Slovakia and Czechia. Slovakia conducted significantly more tests per capita, a difference that became particularly evident around day 128 (January 19, 2021). This discrepancy highlights different approaches to pandemic management: Slovakia adopted a more aggressive testing strategy to identify and isolate cases, while Czechia maintained a more conservative testing protocol. Vertical lines denote the dates of mass testing events in Slovakia.}
	\label{fig:Fig_S4} 
\end{figure}


\begin{figure} 
	\centering
	\includegraphics[width=0.6\textwidth]{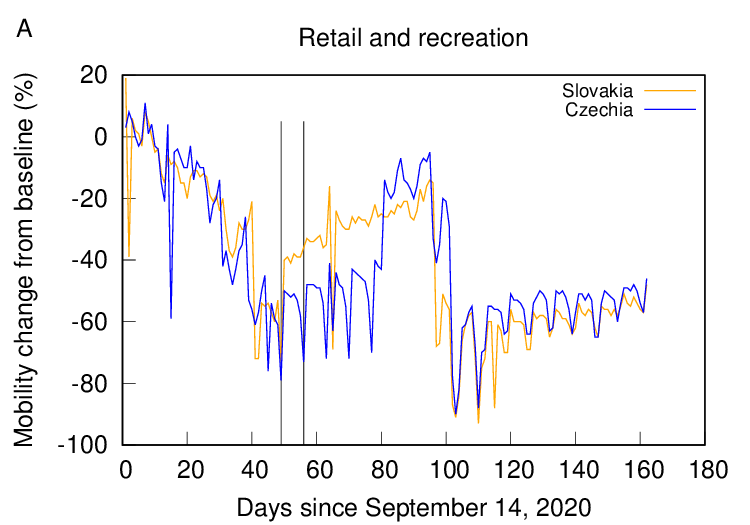}
	\includegraphics[width=0.6\textwidth]{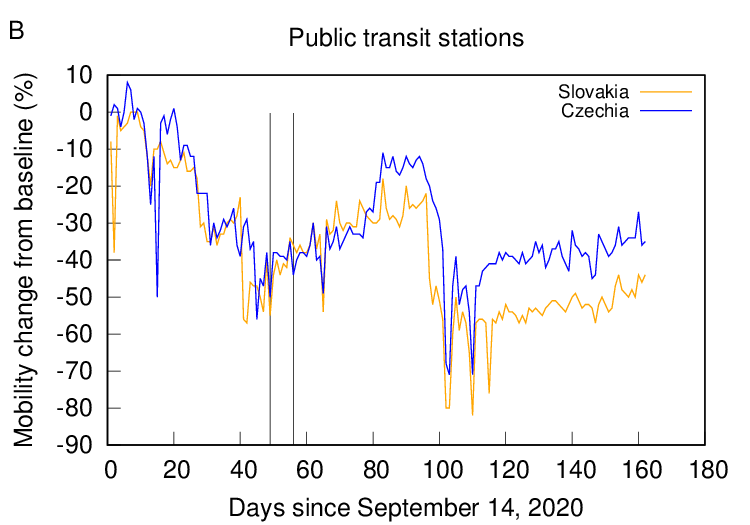}
	\includegraphics[width=0.6\textwidth]{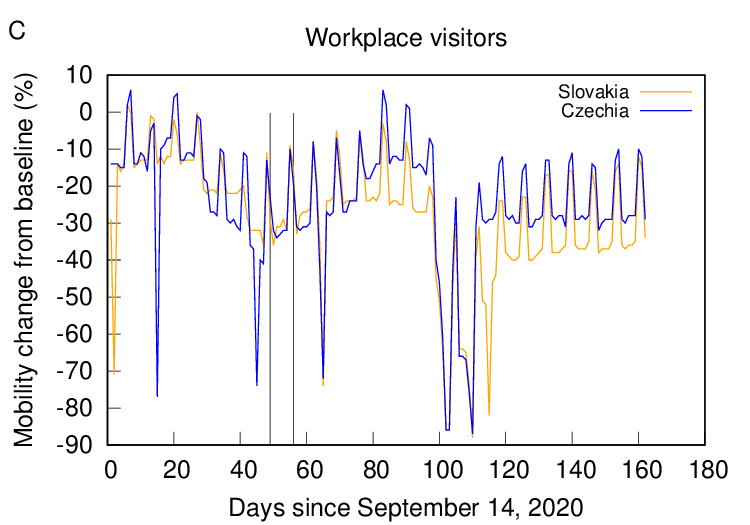} 

	\caption{\textbf{ } caption next page }
	\label{fig:Fig_S5} 
\end{figure}
\addtocounter{figure}{-1}
\begin{figure}[t!] 
  \caption{\textbf{Comparison of mobility trends in Slovakia and Czechia.} The time series display the seven-day rolling average change from the baseline for three categories: retail and recreation (\textbf{A}), public transit stations (\textbf{B}), and workplaces (\textbf{C}). The data span the period from September 14, 2020, to February 22, 2021. A high degree of synchronization in public mobility behaviour between the two countries is evident, both before the mass testing in Slovakia and after day 100. Vertical lines denote the dates of nationwide mass testing events in Slovakia. Data source: Google COVID-19 Community Mobility Reports.}

\end{figure}


\begin{figure} 
	\centering
	\includegraphics[width=0.6\textwidth]{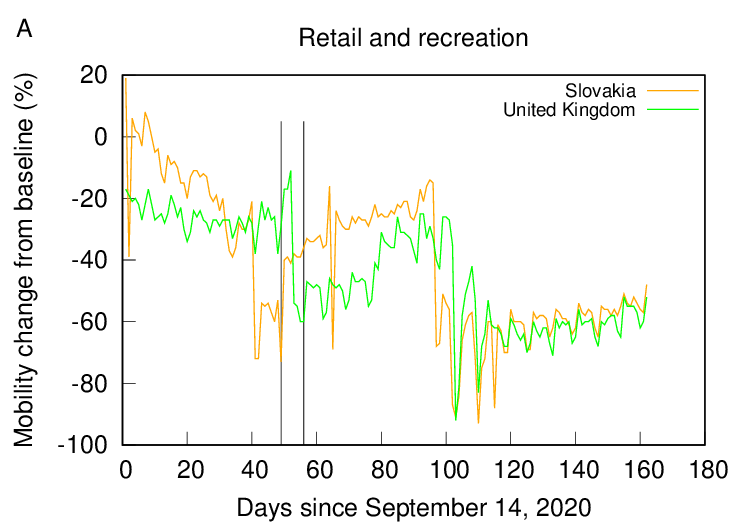}
	\includegraphics[width=0.6\textwidth]{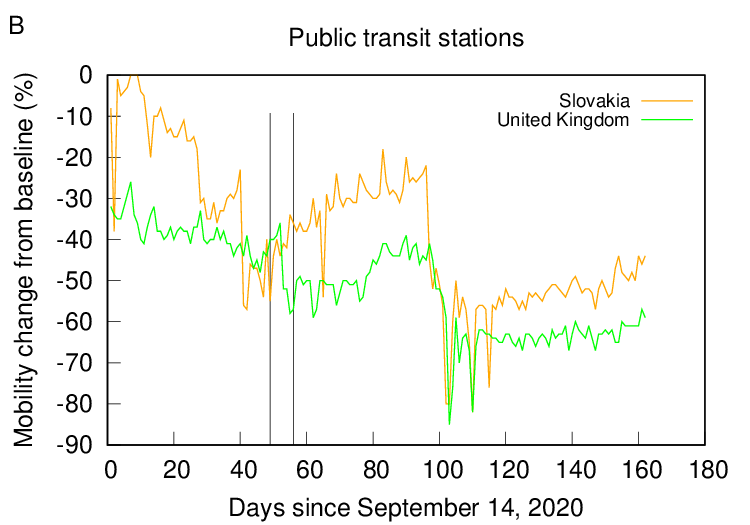}
	\includegraphics[width=0.6\textwidth]{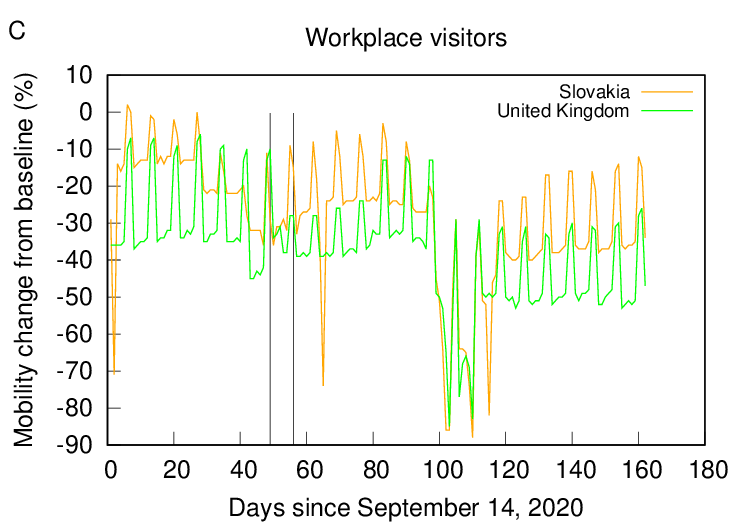} 

	\caption{\textbf{ } caption next page }
	\label{fig:Fig_S6} 
\end{figure}
\addtocounter{figure}{-1}
\begin{figure}[t!] 
\caption{\textbf{Comparison of mobility trends in Slovakia and the United Kingdom.} The time series display the seven-day rolling average change from the baseline for three categories: retail and recreation (\textbf{A}), public transit stations (\textbf{B}), and workplaces (\textbf{C}). The data span the period from September 14, 2020, to February 22, 2021. Vertical lines denote the dates of nationwide mass testing events in Slovakia. Data source: Google COVID-19 Community Mobility Reports.}
\end{figure}


\begin{figure} 
	\centering
	\includegraphics[width=0.9\textwidth]{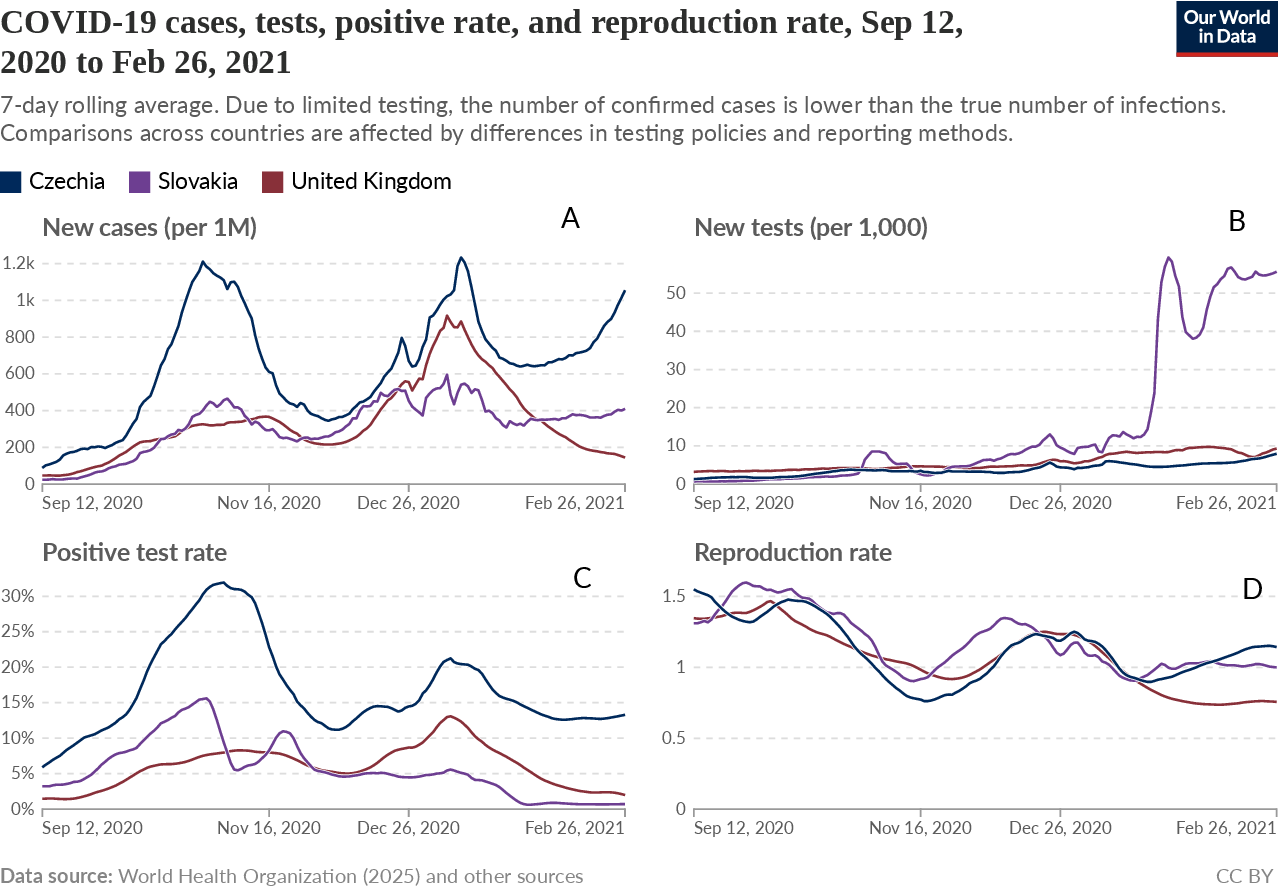} 

	\caption{\textbf{Comparative Dynamics of Key Pandemic Metrics in Slovakia, Czechia, and the United Kingdom (September 2020 – February 2021).} Time series illustrate the seven-day rolling averages for incidence new cases per million, (\textbf{A}), testing intensity new tests per thousand, (\textbf{B}), test positivity rate (\textbf{C}), and the effective reproduction number (\textbf{D}). The comparison highlights divergent policy impacts and epidemic trajectories.
}
	\label{fig:Fig_S7} 
\end{figure}

\begin{table} 
	\centering
	\caption{\textbf{List of countries with a stable effective reproduction number ($R_t$).} The "Days" column shows the duration of the period during which $R_t$ was stable ($R_t = 1 \pm 0.05$). The "New cases" column shows the range of the 7-day rolling average of daily new confirmed COVID-19 cases per million people.}
	\label{tab:sup_rstable} 

	\begin{tabular}{lccccr} 
		\\
		\hline
		Country & Start Date & End Date & Days & New cases & Reference\\
		\hline
		Belgium & Jan 2, 2021 & Feb 18, 2021 & 47 &129--201 &  \cite{Angeli2025}\\
		Brazil & Dec 19, 2020 & Sep 8, 2021 & 263 &100--368 & \\
		Chile & Aug 6, 2020 & Dec 3, 2020 & 119 & 68--100 & \cite{Contreras}\\
		Czechia & Jan 29, 2021 & Feb 9, 2021 & 10 & 641--686 & \\
		Slovakia & Jan 20, 2021 & Mar 3, 2021  & 42 & 308--423 & \\
		Sweden & Apr 27,2020 & May 24,2020  & 27  & 49--60 & \\
		Venezuela & Apr 22, 2021 & Dec 28, 2021 & 250 & 8--63 & \\
		\hline
	\end{tabular}
\end{table}


\clearpage 

\paragraph{Caption for Data S1.}
\textbf{Slovakia's Mass Testing: A Critical Look at the Negative Effects (Figures)}

This repository \cite{Cernak_2026_data} contains the underlying data and high-resolution visualizations supporting the e-letter submitted to Science in response to Pavelka et al. \cite{Pavelka}.

The dataset provides a granular re-analysis of Slovakia's nationwide mass antigen testing campaigns, focusing on the temporal mismatch between testing rounds and the evolution of the effective reproduction number ($R_t$), hospital admissions, and mortality rates.

Key contents of this repository:

1. Figures (PDF/EPS): Detailed visualizations of epidemiological trajectories in Slovakia, the Czech Republic, and the United Kingdom, illustrating the divergence in mobility and clinical outcomes.

2. Processed Data: Aggregated public data from the Google Community Mobility Reports and official national health statistics used to calculate the mortality-to-hospital admission ratios.

3. Analysis Scripts (optional): Gnuplot scripts used to generate the figures, ensuring the reproducibility of our findings.

Our analysis demonstrates that the sustained high mobility and social mixing in Slovakia—driven by the testing policy—contributed to a systemic strain on the healthcare system, resulting in a prolonged mortality plateau during the first quarter of 2021.

Usage: These materials are intended for verification of the arguments presented in the associated e-letter. For inquiries regarding the methodology, please contact the authors through the metadata provided.



\end{document}